\documentclass[aps,pra,twocolumn,superscriptaddress,nofootinbib]{revtex4-2}

\usepackage{graphicx}
\usepackage{latexsym}
\usepackage{amsmath}
\usepackage{amssymb}
\usepackage{amsfonts}
\usepackage{color}
\usepackage{pgf}
\usepackage{amsthm}
\usepackage{dsfont}
\usepackage{tikz}
\usetikzlibrary{matrix}

\usepackage[utf8]{inputenc}
\usepackage[english]{babel}
\usepackage{ulem}
\usepackage{mwe} 
\usepackage{mathtools}
\usepackage{amssymb}
\usepackage{colortbl}

\newcommand{\tr}{\mathrm{Tr}}
\newcommand{\UBS}{\hat{U}_{\mathrm{BS}}}

\begin{document}

\title{Interferometric measurement of the quadrature coherence scale\\ using two replicas of a quantum optical state}

\author{Célia Griffet}
\affiliation{Centre for Quantum Information and Communication, \'Ecole polytechnique de Bruxelles, CP 165, Universit\'e libre de Bruxelles, 1050 Brussels, Belgium}
\author{Matthieu Arnhem}
\affiliation{Centre for Quantum Information and Communication, \'Ecole polytechnique de Bruxelles, CP 165, Universit\'e libre de Bruxelles, 1050 Brussels, Belgium}
\affiliation{Department of Optics, Palacký University, 17. listopadu 12, 771 46 Olomouc, Czech Republic}
\author{Stephan De Bièvre}
\affiliation{Univ. Lille, CNRS, UMR 8524, INRIA - Laboratoire Paul Painlevé, F-59000 Lille, France}
\author{Nicolas J. Cerf}
\affiliation{Centre for Quantum Information and Communication, \'Ecole polytechnique de Bruxelles, CP 165, Universit\'e libre de Bruxelles, 1050 Brussels, Belgium}
\affiliation{James C. Wyant College of Optical Sciences, University of Arizona, Tucson, AZ 85721, USA}

\begin{abstract}
Assessing whether a quantum state $\hat \rho$ is nonclassical (\textit{i.e.}, incompatible with a mixture of coherent states) is a ubiquitous question in quantum optics, yet a nontrivial experimental task because many nonclassicality witnesses are nonlinear in $\hat \rho$. In particular, if we want to witness or measure the nonclassicality of a state by evaluating its quadrature coherence scale, this \textit{a priori} requires full state tomography. Here, we provide an experimental procedure for directly accessing this quantity with a simple linear interferometer involving two replicas (independent and identical copies) of the state $\hat \rho$ supplemented with photon-number-resolving measurements. This finding, which we interpret as an extension of the Hong-Ou-Mandel effect, illustrates the wide applicability of the multicopy interferometric technique in order to circumvent state tomography in quantum optics.
\end{abstract}

\maketitle

\section{Introduction}
In quantum optics, a state $\hat \rho$ is said to be optically classical provided  its Glauber-Sudarshan $P$ function is non-negative \cite{titulaer_correlation_1965}, that is, if there exists a probability density function $P(z)$ such that
\begin{equation}\label{eq:class-state}
\hat \rho=\int P(z) \, |z \rangle\langle z| \, \mathrm{d}^2 z.
\end{equation}
Here, $|z\rangle=\hat D(z)\, |0\rangle$ are the coherent states with $ z\in \mathbb C$, $|0\rangle$ is the vacuum state, and $\hat D(z)=\exp( z\hat a^\dagger-  z^*\hat a)$ is the displacement operator. In other words, an optically classical state is a statistical mixture of coherent states. Detecting the optical nonclassicality of a given state $\hat \rho$ using this definition is, unfortunately, very difficult since the $P$ function can be singular and hence hard to determine both theoretically and experimentally. Further, it is impractical to rule out the existence of a convex mixture of coherent states realizing $\hat \rho$ by checking all realizations of $\hat \rho$. Instead, one is usually forced to resort to so-called witnesses, which only provide a necessary condition on the non-negativity of $P(z)$, and hence a sufficient condition on nonclassicality. Following this line, a large variety of witnesses and measures of optical nonclassicality have been proposed in recent decades in order to detect and quantify nonclassicality; see, \textit{e.g.}, Refs. \cite{hillery_classical_1985,bach_simplex_1986,hillery_nonclassical_1987,hillery_total_1989,lee_measure_1991,agarwal_nonclassical_1992,lee_theorem_1995,lutkenhaus_nonclassical_1995,dodonov_hilbert-schmidt_2000,marian_quantifying_2002,richter_nonclassicality_2002,kenfack_negativity_2004,asboth_computable_2005,ryl_unified_2015,sperling_convex_2015,killoran_converting_2016,alexanian_non-classicality_2018,nair_nonclassical_2017,ryl_quantifying_2017,kwon_nonclassicality_2019,de_bievre_measuring_2019,yadin_general_2016,yadin_operational_2018,luo_quantifying_2019,bohmann_probing_2020}.\par

In this paper, we focus on a witness of optical nonclassicality called the quadrature coherence scale (QCS), introduced in Ref.~\cite{de_bievre_measuring_2019} and further studied in Refs.~\cite{Hoetal19, hertz_quadrature_2020, hertz_relating_2020,hertz_decoherence_2023}. The QCS of the state $\hat \rho$ of $n$ bosonic modes is denoted as $\mathcal C(\hat \rho)$ and its square is defined as \cite{de_bievre_measuring_2019,hertz_quadrature_2020}
\begin{equation}
    \mathcal C^2(\hat \rho)=\frac1{2n\, \mathcal P(\hat \rho)}\left(\sum_{j=1}^{2n}{\tr}\, [\hat \rho,\hat r_j][\hat r_j,\hat \rho]\right),\label{eq:QCSdef} 
\end{equation}
where $\mathbf{\hat{r} }=(\hat{x}_1,\hat{p}_1,\cdots,\hat{x}_n,\hat{p}_n)$ is the vector of the position and momentum quadratures, and $\mathcal P(\hat \rho)= \tr (\hat \rho^2)$ is the purity of the state $\hat \rho$.  
 It was shown in Ref. \cite{de_bievre_measuring_2019} that the QCS is a witness of optical nonclassicality: if $\mathcal C(\hat \rho)>1$, then the state $\hat \rho$ is nonclassical. Since the converse is not true, this witness is not faithful. It nevertheless provides both an upper and lower bound on some suitably defined distance from the set of optically classical states $\cal C_{\text{cl}}$ and, as such, it defines an optical nonclassicality measure \cite{de_bievre_measuring_2019}.  In particular, the larger the QCS, the farther the state is from $\cal C_{\text{cl}}$. In addition, $\mathcal C^2(\hat \rho)$ has a direct physical interpretation as being inversely proportional to the decoherence time of $\hat \rho$~\cite{hertz_quadrature_2020}. The evaluation of the QCS on large families of benchmark states has confirmed its efficiency as an optical nonclassicality measure \cite{de_bievre_measuring_2019,Hoetal19, hertz_quadrature_2020, hertz_relating_2020, hertz_decoherence_2023}.  
\par

The difficulty with measuring the QCS (or many other nonclassicality witnesses) of an arbitrary state $\hat \rho$ arises from 
the fact that it is a nonlinear function of $\hat \rho$, as is obvious from Eq.~\eqref{eq:QCSdef}. At first sight, it thus seems that the task of experimentally accessing $\mathcal C(\hat \rho)$ imposes one to carry out a full tomographic reconstruction of $\hat \rho$ before computing Eq.~\eqref{eq:QCSdef}. Here, we contradict this statement and exhibit an elegant solution that bypasses state tomography and only requires two replicas (independent and identical copies) of $\hat \rho$ in order to measure $\mathcal C(\hat \rho)$. The fact that nonlinear functions of a state may be estimated by quantum interferometry using several replicas of the state has long been known \cite{Ekert-multicopy-method,Moura-multicopy-method}. Specifically, the evaluation of any polynomial in the matrix elements of $\hat \rho$ can be reduced to the measurement of a joint observable over several replicas of $\hat \rho$ (the order of the polynomial translates into the number of replicas) \cite{brun_measuring_2004}. This multicopy interferometric technique has, for example, been used to  witness qubit entanglement by coupling two identical pairs of polarization-entangled photons with beam splitters \cite{Bovino-multicopy}. It was also shown to give direct access to the purity and entanglement of atomic qubits  
in an optical lattice by coupling the atoms pairwise via beam-splitter operations \cite{Moura-multicopy-bosons,Moura-multicopy-bosons-2}. 
Notably, this method has been exploited for experimentally accessing the entanglement in a many-body Bose-Hubbard system \cite{Daley-mulitcopy-bosons,Greiner-Nature}. In particular, the purity of ultracold bosonic atoms in an optical lattice was accessed in Ref. \cite{Greiner-Nature} by probing the average parity of the particle number at the output of a 50:50 beam splitter (realized by controlled tunneling). 

Nonetheless, this multicopy technique has not yet been much exploited in continuous-variable quantum information theory and quantum optics. An exception is the recent derivation of symplectic-invariant entropic uncertainty relations in phase-space by using a multicopy uncertainty observable \cite{hertz_multicopy_2019}. In a related work, it was also shown that nonclassicality witnesses based on matrices of moments of the optical field could be expressed as multicopy observables, resulting in possible experimental schemes to detect nonclassicality \cite{arnhem_multicopy_2022}. Here, we use similar ideas and show that the QCS of a state can be experimentally accessed  by performing an interferometric measurement involving only two replicas of the state impinging on a 50:50 beam splitter. We thereby establish an interesting connection between the nonclassicality of a state and the resulting photon-number distribution at the output of the beam splitter, which can be viewed as an extension of the Hong-Ou-Mandel effect.    

The paper is organized as follows. 
In Sec.~\ref{sec:multicopyPurity}, we describe the aforementioned  two-copy interferometric circuit and show how it enables a measurement of the purity of a single-mode state, which constitutes the denominator of the QCS. 
Then, in Sec.~\ref{sec:multicopyQCS}, we 
show how to use the same measurement to determine the numerator of the QCS, thereby establishing our central result. 
In Sec.~\ref{sec:appdisc}, we illustrate our 
two-copy expression of the QCS [Eq.~\eqref{eq:QCSmulti}]
by applying it to several families of benchmark states. 
As a by-product of our analysis, we provide, in Sec.~\ref{sec:phasespace}, an alternative expression for the QCS that exploits the phase-space formulation of quantum optics and is of interest in its own right.  
In the process, we show that the same interferometric circuit can be used to compute the overlap between two distinct single-mode states. 
For notational convenience, we limit ourselves to the state of a single bosonic mode in Secs. \ref{sec:multicopyPurity}-\ref{sec:phasespace}, but we then show  in Sec.~\ref{sec:multimode} how to extend our results to multimode states. Finally, we conclude in Sec.~\ref{sec:conclusion}.

\section{Two-copy observable for measuring the purity}
\label{sec:multicopyPurity}



As is well known, directly accessing the purity $\mathcal P(\hat{\rho}) =  \tr (\hat{\rho}^2)$ of a quantum state $\hat \rho$ as the expectation value of some observable  is impossible since it is nonlinear in $\hat \rho$. However, as is also well known and readily checked, the purity can be reexpressed as the expectation value of the so-called \textit{swap} operator $\hat S$ taken over two replicas of the state, that is \cite{Ekert-multicopy-method,Moura-multicopy-method},
\begin{equation}\label{eq:multicopyS}
\mathcal P(\hat{\rho}) =  \tr \left((\hat{\rho}\otimes \hat{\rho}) \, \hat S \right),
\end{equation}
where $\hat{S}$ is defined as
$\hat S|\varphi\rangle|\psi\rangle=|\psi\rangle|\varphi\rangle$, $\forall ~|\varphi \rangle,| \psi \rangle$.
Note that $\hat{S}$ is a unitary operator that is also Hermitian, so that it can be viewed as an observable. The fact that we need two replicas here is, of course, simply related to the fact that $\mathcal P(\hat{\rho}) $ is quadratic in $\hat \rho$.

To find the actual measurement scheme, we proceed as in Refs. \cite{Daley-mulitcopy-bosons,Greiner-Nature}. We first write $\hat S$ using the associated mode operators $\hat{a}$ and $\hat{b}$, namely,
\begin{equation}\label{eq:swap}
    \hat{S} = e^{i \frac{\pi}{2} (\hat{a}^\dagger - \hat{b}^\dagger) (\hat{a}- \hat{b})}, 
\end{equation}
as proven in Appendix \ref{sec:appendix-swap-operator}.
In order to measure $\hat S$, we can simply use a 50:50 (balanced) beam splitter. 
The latter corresponds to the Gaussian unitary \cite{WeedbrookRMP}
\begin{equation}  \label{eq:BS50:50}
\hat U_{BS} =e^{\frac{\pi}{4} (\hat{a}^\dagger \hat{b} - \hat{a} \hat{b}^\dagger)} ,
\end{equation}
and, in the Heisenberg picture, it transforms the mode operators $\hat{a}$ and $\hat{b}$ as 
\begin{eqnarray}\label{eq:cd}
    \hat c &\coloneqq& \hat U_{BS}^{\dagger} \, \hat a \, \hat U_{BS}  = (\hat{a} + \hat{b})/\sqrt{2}, \nonumber \\ 
    \hat d &\coloneqq& \hat U_{BS}^{\dagger} \, \hat b \, \hat U_{BS} = (- \hat{a} + \hat{b} )/\sqrt{2},
\end{eqnarray}
where $\hat c$ and $\hat d$ denote the corresponding output mode operators. Hence, Eq. \eqref{eq:swap} can be reexpressed as
\begin{equation}\label{eq:swapC}
    \hat{S} = e^{i \pi \, \hat{d}^\dagger \hat{d}} = (-1)^{\hat{n}_d} ,
\end{equation}
where $\hat{n}_d=\hat{d}^\dagger \hat{d}$ is the photon number in mode $\hat d$.
Here, $\hat d$ stands for the ``difference'' mode, that is, it exhibits destructive interference if we feed the beam splitter with two identical coherent states. From Eq. \eqref{eq:swapC}, it thus appears that the measurement of the purity of a state can be achieved by measuring the average parity of the photon number in the ``difference'' mode at the output of a 50:50 beam splitter after having sent two identical copies of the state at its input. Indeed, Eq.~(\ref{eq:multicopyS}) reduces to
\begin{eqnarray}\label{eq:purity}
\mathcal P(\hat \rho) &=&  \tr \left( (\hat{\rho}\otimes \hat{\rho}) \, (-1)^{\hat n _d} \right),  \nonumber \\
&=&  \tr \left( \UBS (\hat{\rho}\otimes \hat{\rho}) \UBS^\dagger \, (-1)^{\hat n _b} \right) ,
\end{eqnarray}
where the first (second) line corresponds to the Heisenberg (Schr\"odinger) picture. An analogous expression can be found in Refs. \cite{Daley-mulitcopy-bosons,Greiner-Nature}.


\begin{figure}[t]
		\includegraphics[width= 0.8\linewidth]{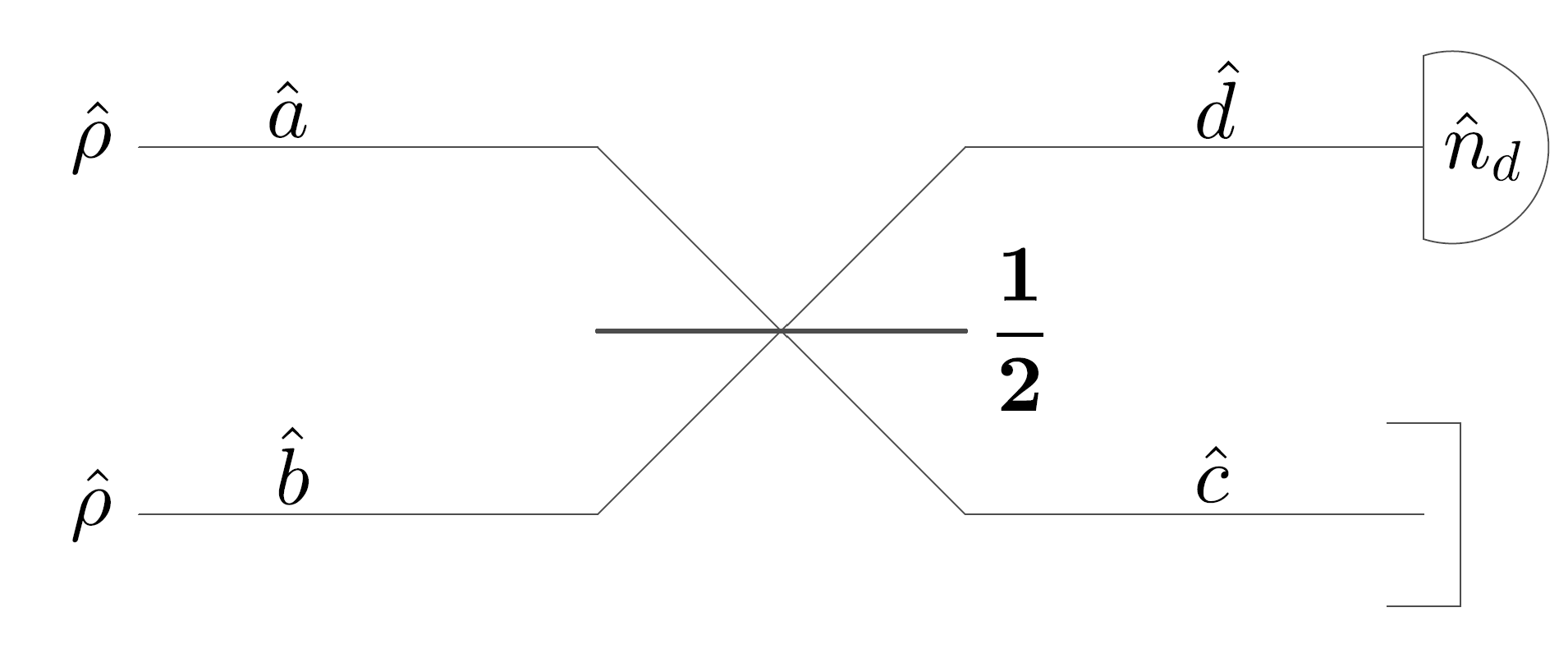}
		\caption{Two-copy circuit implementing the measurement of the purity $\mathcal P(\hat{\rho})$ as well as quadrature coherence scale $\mathcal C(\hat{\rho})$ of a bosonic state $\hat{\rho}$. Two identical copies of the state are sent on a 50:50 beam splitter and the photon number statistics is measured in the output mode $d$ (associated with destructive interference). The mean photon number parity yields the purity, while the numerator and denominator of Eq.~\eqref{eq:QCSmulti} can be accessed separately in order to determine the QCS.}
		\label{Fig:CircuitQCS}
\end{figure}

To sum up, the two-copy interferometric procedure in order to measure the purity of a state $\hat \rho$ is represented in Fig.~\ref{Fig:CircuitQCS}. One must send two identical copies of $\hat\rho$ on a 50:50 beam splitter, which results in the output state $ \UBS \, (\hat \rho\otimes \hat \rho) \, \UBS^\dagger$, and then measure the number of photons with a photon-number resolving detector in the output that is associated with destructive interference. The purity is simply equal to the expectation value of the parity of the photon number in the reduced state $\hat \rho_{d} = \tr_c ( \hat \rho\otimes \hat \rho)$ of mode $\hat d$, namely, 
\begin{equation} \label{eq:purity-bis}
\mathcal P (\hat \rho) = \tr_d \left( \hat \rho_d \,  (-1)^{\hat n_d} \right).
\end{equation}
It makes sense that the output mode $\hat d$ is solely involved here because we expect the purity to be invariant under displacements of $\hat\rho$ in phase space. Indeed, mode $\hat c$ must be disregarded since its state depends on the mean field of the input state $\hat \rho$ (in contrast, the mean field of $\hat \rho_d$ always vanishes, regardless of the mean field of $\hat \rho$). Furthermore, we easily understand that Eq. \eqref{eq:purity-bis} may only depend on the photon number $\hat n_d$ in mode $\hat d$ since the purity is also invariant under rotations of $\hat \rho$ in phase space (remember that a rotation of $\hat \rho$ induces a rotation of $\hat \rho_d$ since the beam splitter unitary $\hat U_{BS}$ is covariant with respect to a pair of identical rotations).

\section{Two-copy observable for measuring the QCS}
\label{sec:multicopyQCS}

The central result of this paper is a two-copy interferometric procedure for measuring the QCS of a state. In the case of a single-mode state $\hat\rho$, Eq. \eqref{eq:QCSdef} reduces to
\begin{equation}\label{eq:QCSsinglemode}
    \mathcal C^2(\hat \rho) = -\, \frac{1}{2 \, \mathcal P(\hat \rho)} \, \tr \left( [\hat{\rho},\hat{x}]^2 + [\hat{\rho},\hat{p}]^2  \right)  ,
\end{equation}
where the purity $\mathcal P(\hat \rho)$ appears in the denominator. The latter can be accessed by applying the multicopy technique as explained in Sec. \ref{sec:multicopyPurity} [see Eq. \eqref{eq:purity-bis}],  so we only need to focus on the numerator $\mathcal N(\hat \rho)$. Writing
\begin{equation}
    \mathcal C^2(\hat \rho) =  \frac{\mathcal N(\hat \rho)}{ \mathcal P(\hat \rho)}   ,
\end{equation}
the numerator can be rewritten as
\begin{eqnarray}\label{eq:QCS2}
\mathcal N(\hat \rho) &=& -\frac{1}{2} \, \tr \left( [\hat{\rho},\hat{x}]^2 + [\hat{\rho},\hat{p}]^2 \right), 
\nonumber \\ 
&=& \frac{1}{2}\left(\int (x-x')^2 \, |\langle x | \hat\rho | x'\rangle|^2 \, \mathrm d x \, \mathrm d x'\right. \nonumber \\ 
&& ~~~~~~~~  + \left.\int (p-p')^2 \, |\langle p | \hat\rho | p'\rangle|^2 \, \mathrm d p \, \mathrm d p' \right), 
\nonumber \\
&=& \frac{1}{2}\left(\int (x-x')^2 \, \langle x,x' | \hat\rho\otimes \hat\rho | x',x\rangle \, \mathrm d x \, \mathrm d x'\right. 
\nonumber \\ 
&& ~~  + \left.\int (p-p')^2 \, \langle p,p' | \hat\rho \otimes \hat\rho | p',p\rangle \, \mathrm d p \, \mathrm d p' \right),
\end{eqnarray}
where we have expanded the first and second terms of the right-hand side in the position and momentum basis, respectively. 

In order to access $\mathcal N(\hat \rho)$ from two copies of $\hat\rho$, we proceed along the same line as for the purity $\mathcal P(\hat \rho)$. It is easy to recognize the two-copy observable $\hat N$ that appears in Eq.~\eqref{eq:QCS2}, that is,
\begin{equation}
  \mathcal N(\hat \rho)   = \tr \left( (\hat{\rho} \otimes \hat{\rho}) \,  \hat N\right) ,
\end{equation}
where 
\begin{equation}
   \hat N = \frac{1}{2} \left[(\hat{x}_a-\hat{x}_b)^2 + (\hat{p}_a-\hat{p}_b)^2\right] \, \hat{S}  ,
\end{equation}
with $\hat{x}_a$ ($\hat p_a$) and $\hat{x}_b$ ($\hat p_b$) denoting the position (momentum) quadratures of modes $\hat a$ and $\hat b$.
The terms involving these quadratures can be simplified by using the quadratures $(\hat{x}_d,\hat{p}_d)$ of the difference mode $\hat{d}$, that is,
\begin{equation}\label{eq:numxandp}
  \hat N =  (\hat{x}_d^2 + \hat{p}_d^2)\, \hat{S} = (1 + 2 \,\hat{n}_d) \, \hat{S}.
\end{equation}
The observable $\hat N$ plays the same role here as the swap operator $\hat S$ in Eq. \eqref{eq:multicopyS}.
Note that the two factors in $\hat N$ are commuting Hermitians, which implies that $\hat N$  is itself Hermitian, as expected. 
Remarkably, we see that $\hat N$ is measurable with the same circuit as the one used for the purity in Fig. \ref{Fig:CircuitQCS}. Indeed, putting together Eqs.~(\ref{eq:swapC}) and (\ref{eq:numxandp}), we obtain the two-copy expression for the QCS,
\begin{equation}\label{eq:QCSmulti}
    \mathcal C^2(\hat \rho) = \frac{\tr_d \left( \hat \rho _d \, (-1)^{\hat{n}_d} \, (1 + 2 \, \hat{n}_d) \right)}{\tr_d \left( \hat \rho_d \, (-1)^{\hat n _d}\right) } ,
\end{equation}
which is our main result. We again need to send two replicas of $\hat \rho$ on a 50:50 beam splitter and measure the photon number $\hat n _d$ of the state $\hat \rho_d$ in the ``difference'' mode $\hat d$. From the measured statistics of $\hat n_d$, we can compute both the numerator and denominator of Eq. \eqref{eq:QCSmulti}.

Similar to before, only the state of mode $\hat d$ is involved in Eq.~\eqref{eq:QCSmulti}, which was expected since the QCS is invariant under displacements in phase space. Further, Eq.~\eqref{eq:QCSmulti} only depends on the photon number $\hat{n}_d$ in mode $\hat d$, which ensures that the QCS is, in addition, invariant under rotations in phase space.

\section{Applications}\label{sec:appdisc}

Let us discuss some applications of this expression of the QCS based on a two-copy observable, as obtained in Eq.~\eqref{eq:QCSmulti}. First, we emphasize that it gives a better grasp on the reason why it detects optical nonclassicality than the one-copy expression given by Eq.~\eqref{eq:QCSsinglemode}. As shown in Appendix~\ref{sec:appendixClassical}, Eq.~\eqref{eq:QCSmulti} can indeed be used to quite easily prove that $\mathcal C^2 \le 1$ for any classical state (mixture of coherent states). In contrast with the proof of Ref.~\cite{de_bievre_measuring_2019}, the physical interpretation is straightforward. Sending two identical coherent states $|\alpha\rangle\otimes|\alpha\rangle$ in the 50:50 beam splitter of Fig.~\ref{Fig:CircuitQCS} results in a coherent state $|\sqrt{2}\, \alpha\rangle$ in mode $\hat c$ and the vacuum state $|0\rangle$  in mode $\hat d$. Hence, the measured value of $\hat n_d$ always vanishes and it is immediate that $\mathcal C^2(|\alpha\rangle\langle\alpha|)=1$ for any coherent state $|\alpha\rangle$. Then, a simple calculation exploiting the Poisson distribution of the photon number in a coherent state is enough to prove that $\mathcal C^2$ can only decrease when mixing coherent states.

Second, we note that the definition of the QCS as given by Eq.~\eqref{eq:QCSdef} [or Eq.~\eqref{eq:QCSsinglemode} for a single mode] is not convenient for computing its value. A number of alternative expressions have been derived in Refs.~\cite{de_bievre_measuring_2019, Hoetal19,hertz_quadrature_2020,hertz_relating_2020} that are more suitable for this purpose in various situations. For example, if the Wigner function of state $\hat \rho$ is known, one can use Eq.~\eqref{eq:QCSWch}, see below. Simple expressions also exist when the state $\hat \rho$ is pure \cite{de_bievre_measuring_2019} or Gaussian \cite{hertz_quadrature_2020,hertz_relating_2020}. In what follows, we analyze the expression given by Eq.~\eqref{eq:QCSmulti} and illustrate its merits. We start by rewriting the expression of the purity, given by Eq.~\eqref{eq:purity}, as
\begin{equation}\label{eq:purity_pnb}
\mathcal P(\hat \rho)
=\sum_{n=0}^\infty p_n \, (-1)^n,
\end{equation}
where 
\begin{equation}\label{eq:pnb}
p_n= \langle n|\hat\rho_{\textrm{out},b}|n\rangle,
\end{equation}
is the probability of finding $n$ photons in the reduced state
\begin{equation}
\hat\rho_{\textrm{out},b}=\tr_a \left(\UBS (\hat{\rho}\otimes \hat{\rho}) \UBS^\dagger \right) ,
\end{equation}
of the measured output mode of the beam splitter. Similarly,
\begin{equation} \label{eq:pnb-num}
  \mathcal N(\hat \rho)   = \sum_{n=0}^\infty p_n \, (-1)^n   \, (1+2n) \,  ,
\end{equation}
so that Eq.~\eqref{eq:QCSmulti} can be rewritten as
\begin{equation}\label{eq:QCS_pn}
{ \mathcal C^2(\hat\rho)=1+2 \, \frac{\sum_n n \, (-1)^n p_n}{\sum_n (-1)^n p_n}.  }
\end{equation}
Whereas it looks simple, Eq.~\eqref{eq:QCS_pn} yields a convenient method to compute the QCS theoretically  only when the $p_n$'s can be easily calculated. Still, the main message of our paper is that, provided the $p_n$'s can be experimentally measured using the above interferometric scheme, we get a direct access to $\mathcal C^2(\hat\rho)$ from Eq.~\eqref{eq:QCS_pn}. 

Interestingly, we may suggestively rewrite $\mathcal C^2(\hat\rho)$ in terms of some peculiar average. Let 
\begin{equation}
\pi_n=\frac{(-1)^n\, p_n}{\sum_n (-1)^n \, p_n},  
\end{equation}
be the quasi-probability distribution associated with $p_n$ (with $\sum_n\pi_n=1$ but $\pi_n \ngeq 0$). Then, Eq.~\eqref{eq:QCS_pn} becomes
\begin{equation}
\mathcal C^2(\hat\rho)=1+2\, \langle n \, \rangle_\pi,  
\end{equation}
where $\langle\cdot\rangle_\pi$ denotes the average with respect to $\pi_n$. Hence, $\langle  n\rangle_\pi>0$ is equivalent to $\mathcal C^2(\hat\rho)>1$ as a sufficient condition for optical nonclassicality. 

At this point, it is instructive to consider what happens when $\hat\rho=|\psi\rangle\langle\psi|$ is a pure state, so that 
\begin{equation}
\mathcal P(|\psi\rangle\langle\psi|)=\sum_n (-1)^n p_n = 1 .
\end{equation}
Since $\sum_n p_n=1$, this implies that only even values of $n$ can be observed at the output of the 50:50 beam splitter ($p_n=0$ if $n$ is odd), which can be viewed as the manifestation of an extended Hong-Ou-Mandel effect. In this case $\pi_n = p_n$ becomes a genuine probability distribution and the QCS is simply given by 
\begin{equation}
\mathcal C^2(|\psi\rangle\langle\psi|) = 1+2\, \bar n  ,  \end{equation}
where $\bar n= \sum_n n \, p_n$ is the average photon number in the measured output mode of the beam splitter. We see here that the QCS always exceeds one unless the pure state $|\psi\rangle$ is a coherent state, in which case $p_{n}=\delta_{n,0}$ and $\bar n=0$. This is consistent with the fact that the only classical pure states are coherent states. The value of the QCS for all other pure states is very easy to find since
$\bar n = \langle \hat a^\dagger \hat a \rangle_\psi - \langle \hat a \rangle_\psi \, \langle \hat a^\dagger \rangle_\psi $, where $\langle \cdot \rangle_\psi = \langle \psi | \cdot | \psi \rangle $ denotes the expectation value in the input pure state $|\psi\rangle$. The number $\bar n$ is sometimes also called the number of thermal (non-coherent) photons of $|\psi\rangle$. Furthermore, it is just equal to the average photon number $\bar n = \langle \hat a^\dagger \hat a \rangle_\psi$ of the input state if the latter state is centered on the origin (if the mean field vanishes). We then immediately recover the known expressions of the QCS for a vacuum squeezed state or a Fock state \cite{de_bievre_measuring_2019}, namely,
\begin{equation}
\mathcal C^2(|\mathrm{Sq}_r\rangle\langle\ \mathrm{Sq}_r |) = \cosh (2r),
\quad
\mathcal C^2(|n\rangle\langle n |) = 1+2n  . 
\end{equation}

\begin{figure}
    \centering
    \includegraphics[width= \linewidth]{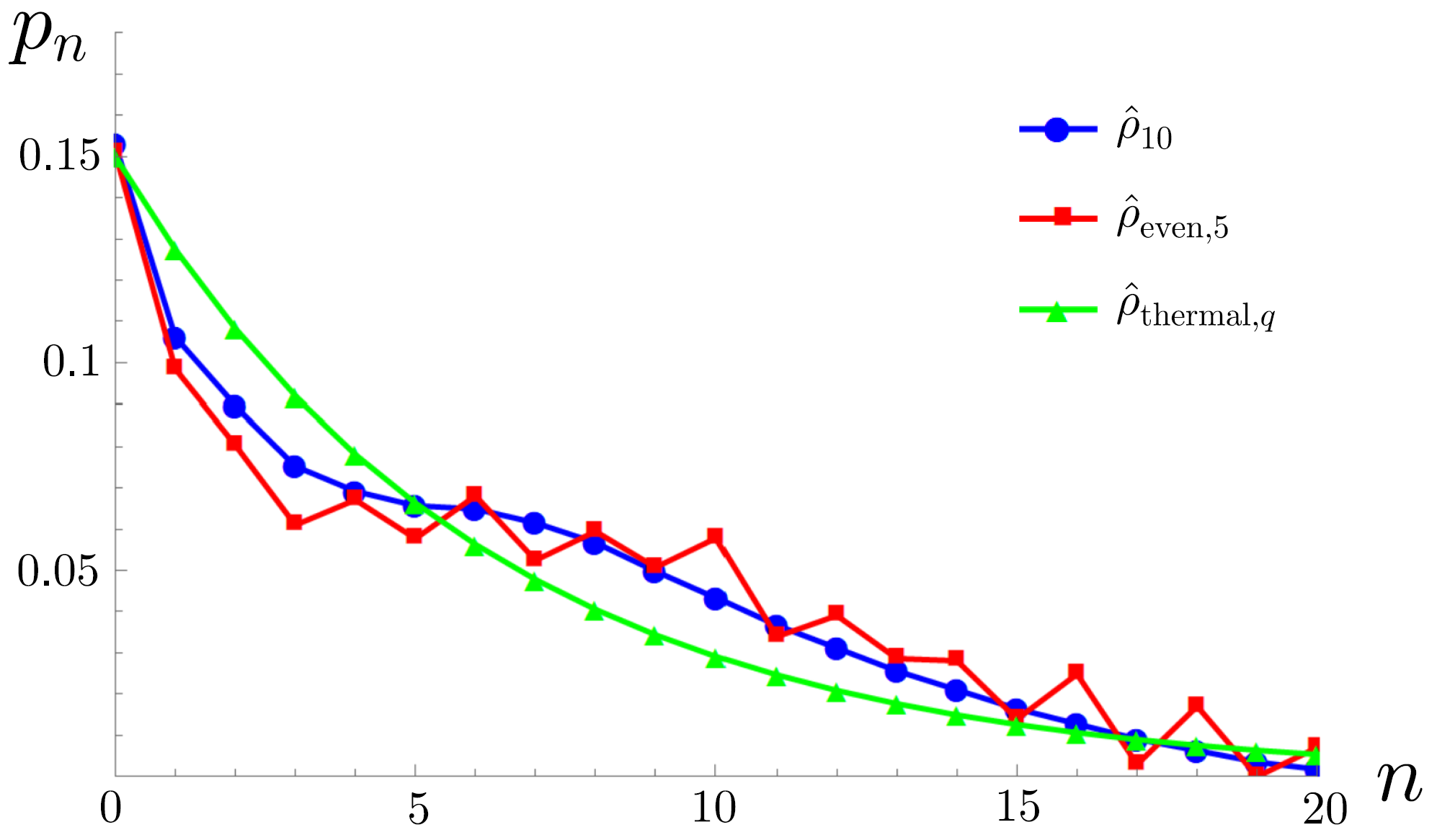}
    \caption{Graphs of the photon number distribution $p_n$ as defined in Eq.~\eqref{eq:pnb} for the states $\hat\rho_{2M}$ and $\hat\rho_{\textrm{even},M}$ with $M=5$, as well as for  $\hat{\rho}_{\mathrm{thermal},q}$ with $q = 0.85$. The value of $q$ is chosen so that $\langle\hat n\rangle_{0.85}=5.7$ is close to the mean photon number in $\hat \rho_{10}$ and $\hat\rho_{\textrm{even},5}$, namely, 5.5 and 6. The purities of these states are given, respectively, by $\mathcal P(\hat\rho_{10})=1/10$, $\mathcal P(\hat\rho_{\textrm{even}, 5})=1/5$, and $\mathcal P(\hat{\rho}_{\mathrm{thermal},q})=0.08$, while their QCS are given by $\mathcal C^2(\hat\rho_{10})=6/5$, $\mathcal C^2(\hat\rho_{\textrm{even}, 5})=13$, and $\mathcal C^2(\hat{\rho}_{\mathrm{thermal},q})=0.08$. }
    \label{fig:pngraphs}
\end{figure}

Let us turn to examples of mixed states $\hat\rho$ and illustrate how the behavior of distribution of the $p_n$'s yields a qualitative idea on the value of the QCS. For this purpose, we consider the following two families of states
\begin{equation}
    \hat\rho_{2M}=\frac1{2M}\sum_{n=1}^{2M} |n\rangle\langle n|, \quad \hat\rho_{\textrm{even},M}=\frac{1}{M}\sum_{n=1}^M | 2n\rangle\langle 2n|.
\end{equation}
with $M\geq 1$. It was shown in Ref.~\cite{de_bievre_measuring_2019} that
\begin{equation}
    \mathcal P(\hat\rho_{2M})=\frac{1}{2M},\quad \mathcal P(\hat\rho_{\textrm{even},M})=\frac1{M},
\end{equation}
and
\begin{equation}
    \mathcal C^2(\hat\rho_{2M})=1+\frac1{M},\quad \mathcal C^2(\hat\rho_{\textrm{even},M})=1+2(M+1). 
\end{equation}
In other words, as $M$ increases, both states are increasingly mixed, but, whereas this goes with a lower nonclassicality for $\hat\rho_{2M}$, the nonclassicality of $\hat\rho_{\textrm{even},M}$ gets higher. As shown in Ref.~\cite{de_bievre_measuring_2019}, this corresponds to increasingly fast oscillations in the Wigner function of $\hat\rho_{\textrm{even},M}$, which are absent for $\hat \rho_{2M}$.  In Fig.~\ref{fig:pngraphs}, we plot the numerically obtained values of $p_n$ for both states for $M=5$. Some details of the underlying computations are provided in Appendix~\ref{sec:appendixpnd}. One observes on these graphs that the $p_n$'s evolve more smoothly as a function of $n$ for $\hat\rho_{10}$ than for $\hat\rho_{\textrm{even}, 5}$. Since for both states
$\mathcal P(\hat\rho)=\sum_{k=0}^{10} (p_{2k}-p_{2k+1})$,
this immediately explains the lower value of the purity for $\hat\rho_{10}$ than for $\hat\rho_{\textrm{even},5}$. In general, it is clear that if the $p_n$'s evolve slowly with $n$, the value of the purity will tend to be smaller. If, on the other hand, they change sharply with successive $n$ (with higher values for even $n$ than for odd $n$), the purity will tend to be larger. This effect is accentuated in the numerator of the QCS because of the extra $n$ factor. This qualitatively  explains the large value of the QCS for $\hat\rho_{\textrm{even},5}$ as a consequence of the sharp variations in the corresponding $p_n$'s as observed in Fig.~\ref{fig:pngraphs}.

To further corroborate this picture, we compute the QCS for thermal states, 
\begin{equation}
\hat\rho_q=(1-q)\sum_n q^n |n\rangle\langle n|,
\qquad 0\leq q<1,
\end{equation}
with a mean photon number $\langle\hat n\rangle_q=q/(1-q)$.
Since sending two identical thermal states at a beam splitter results in the same product of thermal states at the output \cite{WeedbrookRMP}, we simply have $p_n=(1-q) \, q^n$. The graph of $p_n$ for $q=0.85$ is shown in Fig.~\ref{fig:pngraphs} and one sees that it is indeed very smooth as a function of $n$, without fast oscillations as expected since thermal states are well known to be classical. 
Using Eqs.~\eqref{eq:purity_pnb}
and \eqref{eq:pnb-num}, we have
\begin{eqnarray}
    \mathcal P(\hat\rho_q) &=&  (1-q) \sum_{n=0}^\infty (-1)^n q^n = \frac{1-q}{1+q}, \\
    \mathcal N(\hat\rho_q) &=&  (1-q) \sum_{n=0}^\infty (-1)^n q^n (1+ 2n) = \frac{(1-q)^2}{(1+q)^2},
    \hspace{0.5 cm} 
\end{eqnarray}
so that Eq. \eqref{eq:QCS_pn} yields
\begin{equation}
    \mathcal C^2(\hat\rho_q)= \frac{1-q}{1+q}
     = \frac{1}{1+2 \langle\hat n\rangle_q}<1 .
\end{equation}

In short, when the $p_n$'s can be determined, experimentally or otherwise, an inspection of their behavior as a function of $n$ gives a good indication of whether the QCS is large or small. More precisely, one expects that when the $p_n$'s evolve slowly with $n$, the state has a small QCS, whereas sharp variations in the $p_n$'s indicate a large QCS, and hence a nonclassical state. Note that these variations can be viewed as a consequence of quantum  interference, so that we see here again that large interference effects are associated to a large degree of nonclassicality.  When the state is pure (except for coherent states), the full sequence $p_n$ exhibits clear oscillations since all odd terms vanish (extended Hong-Ou-Mandel effect). When the state is mixed, the fluctuations may remain, but are less pronounced.  Thus, overall, Eq.~\eqref{eq:QCS_pn} suggests that the QCS is a measure of the intensity of these fluctuations.


In practice, it must be stressed that the above procedure for measuring the QCS requires optical phase  stability since the two replicas of the state are interfered at a 50:50 beam splitter before being measured. Crucially, the accuracy of the obtained value of the QCS is also conditional on the ability of the detector in Fig.  \ref{Fig:CircuitQCS} to resolve the photon number; this is especially true given that Eq.~\eqref{eq:QCS_pn} depends on the parity of the photon number. In case imperfect photon-number-resolving detectors are used (realized, for example, with arrays of on-off photodetectors) the limitations of the method should be carefully analyzed. Furthermore, the influence of mode mismatch, optical losses, and detector noise should be considered in order to effectively use our protocol (these imperfections could be overcome in an experimental demonstration of our protocol \cite{goldberg_measuring_2023}, see note at the end of this paper).



\section{Phase-space interpretation}\label{sec:phasespace}

Combining the results of the previous sections with  phase-space formalism, we can express the purity and QCS of a state $\hat \rho$ in terms of the Wigner function of the output state $\hat \rho_d$ that is found in the output mode $\hat d$ (associated with destructive interference). This in turn yields expressions of the purity and QCS in terms of the Wigner function  of state $\hat \rho$. We refer to Appendix~\ref{sec:appendixWigner} for the basics of Wigner functions and the conventions we use here.

It is instructive to consider the purity as a special case of the overlap of two input states $\hat\rho_a$ and $\hat\rho_b$ impinging on the 50:50 beam splitter in Fig. \ref{Fig:CircuitQCS}.
If their respective Wigner functions are denoted by $W_a(x,p)$ and $W_b(x,p)$, then the Wigner function of the state on mode $\hat d$,
\begin{equation}
    \hat \rho_{d} = \tr_c ( \hat \rho_a \otimes \hat \rho_b) = \tr_a \left(\UBS (\hat{\rho}_a\otimes \hat{\rho}_b) \UBS^\dagger \right),
\end{equation}
is given by the (scaled) convolution \cite{Ulf_Leonhardt_2003} 
\begin{equation}\label{eq:beamsplitWd}
W_d(x,p)=2 \! \int \! W_a(x',p') \, W_b(x'+\sqrt{2}x,p'+\sqrt{2}p)  \,\mathrm{d}x' \, \mathrm{d}p'.
\end{equation}
Its value at the origin in phase-space is thus
\begin{eqnarray}\label{eq:overlapwigner}
W_d(0,0) &=& 2 \int W_a(x',p') \, W_b(x',p') \,dx' \, dp'.
\end{eqnarray}

Hence, using the overlap formula recalled in Appendix~\ref{sec:appendixWigner}, given by Eq. \eqref{eq:overlapformula}, we have\footnote{Note that $W_d(x,p)\ge 0$, $\forall x,p$, at the output of a 50:50 beam splitter with arbitrary input states $\hat\rho_a$ and $\hat\rho_b$ \cite{zach}, which is consistent with the fact that the overlap $\tr(\hat\rho_a \, \hat\rho_b) $ is non-negative.}
\begin{equation}\label{eq:overlapWd}
  \tr (\hat\rho_a \, \hat\rho_b) = \pi \, W_d(0,0) .
\end{equation}
Then, using the well-known property that the value of a Wigner function evaluated at the origin is proportional to the expectation value of the photon number parity, 
[see Appendix \ref{sec:appendixWigner}, Eq.~\eqref{eq:parity-wigner-at-origin}], we conclude that
\cite{Daley-mulitcopy-bosons,Greiner-Nature}
\begin{eqnarray}  \label{eq:overlap-parity-n_d}
\tr (\hat\rho_a \, \hat\rho_b) =\pi \, W_d(0,0) = \tr_d \left(\hat\rho_d \, (-1)^{\hat{n}_d} \right) .
\end{eqnarray}
This implies that the overlap between states 
$\hat\rho_a$ and $\hat\rho_b$ can be accessed by measuring the expectation value of the photon number parity on the output mode $\hat d$ (associated with destructive interference) of a 50:50 beam splitter using the scheme of Fig. \ref{Fig:CircuitQCS} but with input states $\hat\rho_a$ and $\hat\rho_b$. 
Of course, the purity corresponds to the special case where $\hat\rho_a$ and $\hat\rho_b$ are both equal to $\hat \rho$ in which case, Eq.~\eqref{eq:overlap-parity-n_d} reduces to Eq.~\eqref{eq:purity-bis}.
We will now use Eq.~\eqref{eq:QCSmulti} to express the QCS in terms of the Wigner function of the state $\hat \rho_d $ and its derivatives evaluated at the origin. For the denominator, the desired expression results from Eq. \eqref{eq:overlap-parity-n_d} where $\hat \rho_a = \hat \rho_b$, that is,
\begin{eqnarray}
\mathcal P(\hat \rho)   &=& \tr \left[ (\hat{\rho} \otimes \hat{\rho}) \, \hat S \right], \nonumber\\
 &=& \tr_d \left[ \hat \rho_d \,  (-1)^{\hat n_d} \right],
 \nonumber \\
 &=& \pi \, W_d(0,0) .
\end{eqnarray}
For the numerator $\mathcal N(\hat \rho)$, we can write 
\begin{eqnarray}
\label{eq-numerator-Wd}
\mathcal N(\hat \rho)   &=& \tr \left[ (\hat{\rho} \otimes \hat{\rho}) \, \hat N \right], \nonumber\\
 &=& \tr_d \left[ \hat \rho_d  \, (\hat{x}_d^2 + \hat{p}_d^2)\, (-1)^{\hat n_d} \right],
 \nonumber\\
 &=&  - {\pi \over 4} \Delta W_d\bigg|_{x=0,p=0},
\end{eqnarray}
where $\Delta$ stands for the Laplacian. The last equality in Eq. \eqref{eq-numerator-Wd} is obtained by again using the overlap formula as well as the Weyl transform of the operator $({\hat x}^2_d + {\hat p}^2_d) \, (-1)^{\hat n_d} /\pi$ 
[see Appendix \ref{sec:appendixWigner}, Eq.~\eqref{eq:laplacian-wigner-at-origin}]. As a result, we obtain
\begin{eqnarray}
\mathcal C^2(\hat \rho)  
= \frac{\mathcal N (\hat \rho)}{\mathcal P (\hat \rho)}=  - {1\over 4} {\Delta W_d \over W_d}\bigg|_{x=0,p=0}.
\label{eq:QCSDeltaW}
\end{eqnarray}

Finally, we can express $\mathcal C^2(\hat \rho)$ in terms of the Wigner function of the state $\hat \rho$ itself (instead of $\hat \rho_d$). 
From Eq.~\eqref{eq:beamsplitWd} with $W_a=W_b=W$ and partial integration, one readily sees that
\begin{eqnarray}
\Delta W_d(0,0)&=& 4 \! \int \! W(x',p') \, \Delta W(x',p')  \,\mathrm{d}x' \, \mathrm{d}p', \\
&=& -\|\nabla_\alpha W\|^2.
\end{eqnarray}
Here, $\|\cdot\|_2$ stands for the $L^2$-norm, meaning for example $\|W\|_2^2:=\int|W(\alpha)|^2 \, \mathrm d^2 \alpha$ and $\nabla_\alpha=(\partial_{\alpha_1},\partial_{\alpha_2})$. Using this together with  Eq.~\eqref{eq:overlapwigner}  for $W_a=W_b=W$, Eq.~\eqref{eq:QCSDeltaW} can be reexpressed as 
\begin{equation}\label{eq:QCSWch}
\mathcal C^2(\hat\rho)=\frac14\frac{\|\nabla_\alpha W\|_2^2}{\|W\|_2^2},
\end{equation}
which is a formula originally derived in Ref. \cite{de_bievre_measuring_2019}. 





Using these results, one can easily recover the well-known formula for Gaussian purity, overlap and QCS (see Appendix \ref{sec:appendixGaussian}).

\section{Multimode case}
\label{sec:multimode}
\begin{figure}[t]
	\begin{center}
		\includegraphics[width= 0.8\linewidth]{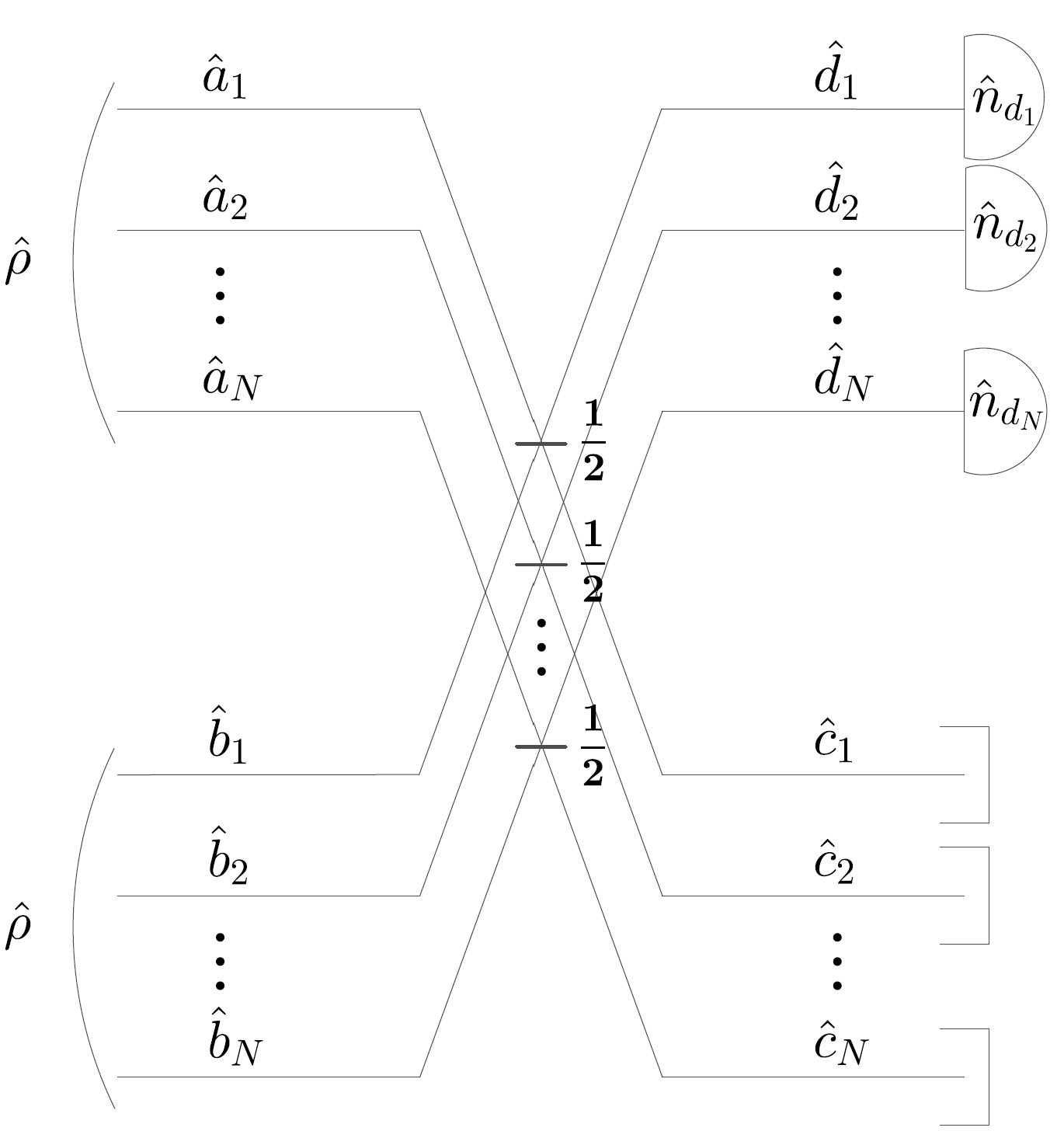}
		\caption{Circuit implementing the measurement of the purity and QCS for a $N$-mode state $\hat \rho$. Here, two identical copies of the state are sent in a stack of 50:50 beam splitters and the photon number statistics is measured in the output modes $\hat d_k$, with $k=1,\cdots N$.}
		\label{fig:Circuitmultimode}
	\end{center}
\end{figure}

The extension of the multicopy interferometric method to the measurement of the QCS and purity of a $n$-mode state is immediate. Note first that the swap operator can be written as
\begin{equation}
    \hat{S} = e^{i \frac{\pi}{2} \sum_{k=1}^N (\hat{a}_k^\dagger - \hat{b}_k^\dagger)(\hat{a}_k - \hat{b}_k)},
\end{equation}
where $k$ is the mode index. Coupling the set of $\hat a$ and $\hat b$ modes pairwise is done with a stack of $N$ beam splitters (see Fig.~\ref{fig:Circuitmultimode}), each of them effecting the unitary
\begin{equation}
\hat U_{BS_k} =e^{\frac{\pi}{4} (\hat{a}^\dagger_k \hat{b}_k - \hat{a}_k \hat{b}^\dagger_k)}.
\end{equation}
Defining $\hat c _k$ and $\hat d_k$ as in Eq. \eqref{eq:cd}, one finds: 
\begin{equation} 
\hat S = \prod_{k=1}^N e^{i \pi \hat{d}^\dagger_k \hat{d}_k} = (-1)^{\sum_{k=1}^N \hat{n}_{d_k}}.
\end{equation} 
Hence, the multimode purity is expressible as before,

\begin{equation}
    \mathcal P(\hat \rho) =\tr \left( (\hat \rho\otimes \hat \rho) \hat S \right)= \tr \left(\hat \rho_d (-1)^{\sum_k \hat{n}_{d_k} }\right),
\end{equation}
where 
$$
\hat \rho_d=\tr_c (\rho\otimes\rho).
$$
Again, the same circuit can be used to measure the multimode version of the QCS with
\begin{equation}\label{eq:QCSmultimode}
   \mathcal C^2(\hat \rho) = \frac{\tr \left(\hat \rho _d \sum_{k'} (-1)^{\sum_k \hat{n}_{d_k}}( 1 + 2 \hat{n}_{d_{k'}})\right)}{n \ \tr \left(\hat \rho_d (-1)^{\sum_k \hat n _{d_k}}\right)}.
\end{equation}

\section{Conclusion}
\label{sec:conclusion}
The quadrature coherence scale is an efficient nonclassicality  measure, which can be expressed through several equivalent formulas, making  it relatively easy to compute for a large variety of states \cite{de_bievre_measuring_2019,Hoetal19,hertz_quadrature_2020,hertz_relating_2020,hertz_decoherence_2023}. It has a clear physical interpretation, notably because it is inversely proportional to the decoherence time of the state \cite{hertz_quadrature_2020}. However, since it is a nonlinear function of the density matrix, its measurement would \textit{a priori} seem to require a complete quantum state tomography. We have shown that this problem can be avoided through the use of a simple two-copy interferometric measurement scheme using a 50:50 beam splitter associated with photon counting. The method can easily be adapted to multimode systems, in which case one needs to couple the modes pairwise using a stack of 50:50 beam splitters.
The challenge of this procedure is of course the need to ensure the interferometric stability of the joint measurement of two replicas together with reasonably low mode-mismatch, optical losses in the circuit, and extra noise, as well as the need for photon-number resolving detectors. 

The underlying multicopy technique used here was put forward in the 2000s \cite{brun_measuring_2004,Ekert-multicopy-method,Moura-multicopy-method} and was more recently applied to bosonic atoms in optical lattices \cite{Daley-mulitcopy-bosons,Greiner-Nature} as well as continuous-variable quantum optical systems \cite{hertz_multicopy_2019,arnhem_multicopy_2022}. 
The present work further extends the range of applicability  of this technique in quantum optics.

\bigskip
\noindent \textit{Note:} Recently, our protocol for measuring the QCS has been successfully implemented in an experiment involving state-of-the-art (superconducting) photon-number-resolving detectors \cite{goldberg_measuring_2023}.

\acknowledgments
The authors thank Z. Van Herstraeten for very useful comments. C.G. is Research Fellow of the Fonds de la Recherche Scientifique-FNRS. M.A.  acknowledges support from the European Union's HORIZON Research and Innovation Actions under Grant Agreement No. 101080173 (CLUSTEC) and European Union’s 2020 research and innovation programme (CSA - Coordination and support action, H2020-WIDESPREAD-2020-5) under Grant Agreement No. 951737 (NONGAUSS) and support from MEYS Czech Republic and the European Union’s Horizon 2020 (2014-2020) research and innovation framework programme under Grant No. 731473 (Project No. 8C20002 ShoQC). Project ShoQC has received funding from the QuantERA ERA-NET Cofund in Quantum Technologies implemented within the European Union’s Horizon 2020 program. S.D.B. acknowledges support by the Agence Nationale de la Recherche under Grant No. ANR-11-LABX-0007-01 (Labex CEMPI) and by the Nord-Pas de Calais Regional Council and the European Regional Development Fund through the Contrat de Projets \'Etat-R\'egion (CPER). N.J.C. acknowledges support by the Fonds de la Recherche Scientifique – FNRS under Grant No T.0224.18 and by the European Union under Project ShoQC within ERA-NET Cofund in Quantum Technologies (QuantERA) program.

\bigskip

\appendix

\section{Swap operator $\hat{S}$ in Fock space}
\label{sec:appendix-swap-operator}


The swap operator $\hat S $ naturally extends to the infinite-dimensional Fock space of a bosonic mode (or harmonic oscillator), in which case it is convenient to express it in terms of mode operators $\hat{a}$ and $\hat{b}$. For this purpose, we introduce the Hamiltonian $\hat H = -(\hat{a}^\dagger - \hat{b}^\dagger) (\hat{a}- \hat{b})$ and write the Heisenberg evolution of $\hat a$ and $\hat b$ (setting $\hbar=1$), namely, 
\begin{eqnarray}
    \hat{a}(t) &=& \exp(it\hat{H})\, \hat{a} \, \exp(-it\hat{H}) , \nonumber \\
    \hat{b}(t) &=& \exp(it\hat{H})\, \hat{b} \, \exp(-it\hat{H}).
\end{eqnarray}
Hence,
\begin{eqnarray}
    i \, \frac{d \hat{a}(t)}{dt} &=& [\hat{a}(t),\hat H] = - \hat{a}(t) + \hat{b}(t),  \nonumber \\
    i \, \frac{d \hat{b}(t)}{dt} &=& [\hat{b}(t),\hat H] = - \hat{b}(t) + \hat{a}(t),
\end{eqnarray}
resulting in the solution
\begin{eqnarray}
     \hat{a}(t) &=&  \frac{1+e^{2it}}{2}\, \hat{a} + \frac{1-e^{2it}}{2}\, \hat{b},  \nonumber \\
     \hat{b}(t) &=& \frac{1-e^{2it}}{2}\, \hat{a} + \frac{1+e^{2it}}{2}\, \hat{b}.
\end{eqnarray}
Setting $t=\pi/2$, we obtain
\begin{eqnarray}\label{eq:swapAB}
     \hat{a}(\pi/2) =   \hat{b},  \qquad
     \hat{b}(\pi/2) =  \hat{a} ,
\end{eqnarray}
effecting a swap of the two modes. This implies that $\hat S = e^{-i \frac{\pi}{2} \hat H}$, which proves Eq. \eqref{eq:swap}.
$\hfill\square$

Using the unitary $\hat U_{BS}$ corresponding to a 50:50 beam splitter as defined in Eq. \eqref{eq:BS50:50}, we can express the swap operator \eqref{eq:swap} in the Heisenberg picture as
\begin{equation}
   \hat S = \hat U_{BS}^\dagger \, e^{i \pi \, \hat{b}^\dagger \hat{b}}  \, \hat U_{BS}  .
\end{equation}
Hence, the swap operator can be implemented by processing the two modes (forwards) through a 50:50 beam splitter, acting with a $\pi$-phase shift in the second mode, and then processing the two modes (backwards) again through a 50:50 beam splitter. As we could expect, $\hat S$ is simply the Gaussian unitary that corresponds to a Mach-Zehnder interferometer with a $\pi$-phase in one of the two arms, effecting a swap between the two modes.

As an illustration, let us check the effect of $\hat S $ on two coherent states $|\alpha\rangle$ and $|\beta\rangle$. We have
\begin{eqnarray}
    \hat{S} \, |\alpha\rangle \otimes |\beta\rangle &=& \hat U_{BS}^\dagger \, e^{i \pi \, \hat{b}^\dagger \hat{b}}  \, \hat U_{BS} ~ |\alpha\rangle \otimes |\beta\rangle, \nonumber \\ 
    &=& \hat U_{BS}^\dagger \, e^{i \pi \, \hat{b}^\dagger \hat{b}}  ~  \Big|\frac{\alpha+\beta}{\sqrt{2}}\Big\rangle \Big|\frac{-\alpha+\beta}{\sqrt{2}}\Big\rangle,    \nonumber \\
    &=& \hat U_{BS}^\dagger ~    \Big|\frac{\alpha+\beta}{\sqrt{2}}\Big\rangle \Big|\frac{\alpha-\beta}{\sqrt{2}}\Big\rangle,    \nonumber \\
    &=&    \Big|\frac{(\alpha+\beta)-(\alpha-\beta)}{2}\Big\rangle \Big|\frac{(\alpha+\beta)+(\alpha-\beta)}{2}\Big\rangle,    \nonumber \\
    &=& |\beta\rangle \otimes |\alpha\rangle,
\end{eqnarray}
where we have used the fact that a product of coherent states $|\alpha \rangle \otimes |\beta \rangle $ results under $\hat U_{BS}$ into another product of coherent states 
$|(\alpha+\beta)/\sqrt{2} \rangle \otimes |(-\alpha+\beta)/\sqrt{2} \rangle$. Alternatively, we can simply check that $ \hat{S}\,|\alpha\rangle \otimes |\beta\rangle$ is a common eigenstate of $\hat{a}$ and $\hat{b}$ with respective eigenvalues $\beta$ and $\alpha$ (note the interchange). Indeed, we have
\begin{eqnarray}
\hat{a} \, ( \hat{S} \, |\alpha\rangle \otimes |\beta\rangle) &=& \hat{S}\, (\hat{S}^\dagger \, \hat{a} \, \hat{S}) \, |\alpha\rangle \otimes |\beta\rangle, \nonumber \\
&=& \hat{S}\, \hat{b} \, |\alpha\rangle \otimes |\beta\rangle, \nonumber \\
&=& \beta \, (\hat{S}\, |\alpha\rangle \otimes |\beta\rangle),
\end{eqnarray}
and a similar equation holds for $\hat{b} \, ( \hat{S} \, |\alpha\rangle \otimes |\beta\rangle)$.



 



\section{Proof that QCS $\le 1$ for classical states}
\label{sec:appendixClassical}

It was proven in Ref.~\cite{de_bievre_measuring_2019} that the QCS is smaller than or equal to $1$ for all classical states, namely mixtures of coherent states $\hat{\rho} = \int P(\alpha) \, |\alpha \rangle \langle \alpha | \, \mathrm{d}^2\alpha$.
Here, we provide an alternative (much simpler) proof of this result by taking advantage of the twocopy expression for the QCS given by Eq.~\eqref{eq:QCSmulti}.

To measure the QCS, we inject two identical copies of state $\hat{\rho}$ in the circuit of Fig. \ref{Fig:CircuitQCS},
so the input state is
\begin{equation}
    \hat{\rho} \otimes \hat{\rho}  = \int P(\alpha) \, P(\beta) \, |\alpha \rangle \langle \alpha |  \otimes |\beta \rangle \langle \beta | \, \mathrm{d}^2\alpha \, \mathrm{d}^2\beta
\end{equation}
with $\int P(\alpha) \, \mathrm{d}^2\alpha=1$ and $P(\alpha)\ge 0, \forall \alpha$. Each product term  $|\alpha \rangle \otimes |\beta \rangle $ of this mixture results, at the output of the 50:50 beam splitter, into another product of coherent states 
$|(\alpha+\beta)/\sqrt{2} \rangle \otimes |(-\alpha+\beta)/\sqrt{2} \rangle$. We only care here about the reduced output state in mode $\hat d$, that is
\begin{equation}
    \hat\rho_d =  \int P_d(\gamma) \, |\gamma \rangle \langle \gamma | \,  \mathrm{d}^2\gamma ,
\end{equation}
where we have made the change of variables $\delta = (\alpha+\beta)/\sqrt{2}$ and $\gamma = (-\alpha+\beta)/\sqrt{2}$, and then have integrated over variable $\delta$. Here, $P_d(\gamma)$  of course depends on $P(\alpha)$ but its explicit expression is irrelevant for the proof; we only note that $P_d(\gamma)\ge 0$, $\forall \gamma$, since it is a probability density, so that $\hat\rho_d$ is a classical state too. 



We are left with having to compute the mean values of $(-1)^{\hat n_d}$ and $(-1)^{\hat n_d} (1+ 2 \hat n_d)$ based on the distribution of the photon number $n_d$ in state $\hat\rho_d$. Using the probability distribution of the photon number in a coherent state $|\gamma \rangle$,
\begin{equation}
    p_{\gamma}(n) = e^{-|\gamma|^2} \frac{|\gamma|^{2n}}{n!},
\end{equation}
we obtain the following expressions
\begin{eqnarray}
    \sum_{n_d=0}^{\infty} (-1)^{n_d} \, p_{\gamma}(n_d) &=& e^{-| \gamma |^2 } , \\
    \sum_{n_d=0}^{\infty} (-1)^{n_d} \, (1+ 2 n_d) \, p_{\gamma}(n_d) &=& e^{-| \gamma |^2 }(1- |\gamma |^2 ) .
\end{eqnarray}
Hence, taking the average over $\gamma$, we obtain the simple expression for the QCS:
\begin{align}
    \mathcal C^2(\rho) &= \frac{
    \int P_d(\gamma) \, e^{-|\gamma|^2} (1-|\gamma|^2)  \, \mathrm{d}^2\gamma}
    {\int P_d(\gamma) \, e^{-|\gamma|^2} \,  \mathrm{d}^2\gamma} , \\
    &=1-\frac{
    \int P_d(\gamma) \, e^{-|\gamma|^2} |\gamma|^2 \,  \mathrm{d}^2\gamma}
    {\int P_d(\gamma) \, e^{-|\gamma|^2} \,  \mathrm{d}^2\gamma}.
\end{align}
Since the second term in this expression is always positive, the QCS can only be smaller than or equal to $1$ for classical states.
$\hfill \square$

\section{Computation of the $p_n$ for phase-invariant states}\label{sec:appendixpnd}
In this appendix we briefly indicate how to compute the $p_n$ in Sec.~\ref{sec:appdisc} for phase-invariant states.

First, let us derive the photon number probability at the output if we put $N$ photons in mode $\hat a$ and $N'$ photons in mode $\hat b$. Then the initial state is given by:
\begin{equation}
| N N' \rangle_{a,b} = \frac{\hat a ^{\dagger N}}{\sqrt{N !}}\frac{\hat b ^{\dagger N'}}{\sqrt{N '!}} | 00 \rangle_{a,b}.
\end{equation}
Expanding this state in the number bases associated to the $c$ and $d$ modes yields:
\begin{align}
   | N N' \rangle_{a,b} &= \frac{(-\hat c^\dagger + \hat d^\dagger) ^N}{\sqrt{2^N N !}}\frac{ (\hat c^\dagger + \hat d^\dagger) ^{N'}}{\sqrt{2^{N'} N '!}} | 00 \rangle_{a,b},\\
   &= \sum_{n=0}^{N+N'}\sum_{n'=0}^{N'} c(n,n',N,N') |N+N'-n,n \rangle_{c,d},
\end{align}
where 
\begin{equation}
    c(n,n',N,N') = \frac{(-1)^{N-n+n'} \sqrt{N! N'! (N+N'-n)! n!}}{\sqrt{2^{N+N'}}(n-n')!n'!(N-n+n')!(N'-n')!}.
\end{equation}
Consequently,  the photon probability distribution in mode $\hat d$ is equal to:
\begin{equation}
    p_n^{N,N'} = \Big(\sum_{n'=0}^{N'}c(n,n',N,N')\Big)^2.
\end{equation}

If the state is $\hat \rho = \sum_m \lambda_m |m\rangle \langle m|$, then, as the input state, we have:
\begin{equation}
   \hat \rho \otimes \hat \rho =  \sum_m \sum_{m'} \lambda_m \lambda_{m'} |m\rangle \langle m| \otimes |m'\rangle \langle m'|,
\end{equation}
 and the probability distribution is given by: 
\begin{equation}
    p_n = \sum_m \sum_{m'} \lambda_m \lambda_{m'} p_n^{m,m'}.
\end{equation}
These expressions are readily evaluated numerically and were used to produce the plots in Fig.~\ref{fig:pngraphs}. It is also clear that these expressions are not a convenient starting point to analytically compute the QCS of the state $\hat \rho$. 

\section{Useful properties of Wigner functions}
\label{sec:appendixWigner}


Setting $\hbar=1$, the Weyl transform $\tilde A (x,p)$ of a linear operator $\hat A$ is defined as
\begin{equation}
\tilde A(x,p) = {1\over 2\pi} \int \langle x- {y\over 2} | \hat A | x + {y\over 2} \rangle \, e^{ipy} \, dy.
\end{equation}
Applied to a density operator $\hat \rho$, the Weyl transform gives the Wigner function 
\begin{equation}\label{eq:wignerdef1}
W(x,p) = {1\over 2\pi} \int \langle x- {y\over 2} | \hat \rho | x + {y\over 2} \rangle \, e^{ipy} \, dy,
\end{equation}
while the Weyl transform of the identity operator $\hat \openone$ is simply equal to the constant function $1/2\pi$. Note also that the Weyl transforms of operators $f(\hat x)$ and $g(\hat p)$ are $f(x)/2\pi$ and $g(p)/2\pi$, respectively, where $f$ and $g$ are arbitrary functions. For any two linear operators $\hat A_1$ and $\hat A_2$, the overlap formula reads
\begin{equation} \label{eq:overlapformula}
\tr (\hat A_1 \,\hat A_2)
 =  2 \pi \int \tilde A_1(x,p) \, \tilde A_2(x,p) \, dx \, dp,
\end{equation}
which implies, for example, that
\begin{equation} 
\tr(\hat \rho)
 =  \int W(x,p) \, dx \, dp .
\end{equation}
In order to link the purity with the Wigner function of the reduced state $\hat \rho_d$ in the ``difference'' mode (see Sec.~\ref{sec:phasespace}), we need to compute the Weyl transforms of the parity operator $(-1)^{\hat n}/\pi$, namely

\begin{eqnarray}
\lefteqn{ {1\over 2\pi} \int \langle x- {y\over 2} | {(-1)^{\hat n} \over \pi}| x + {y\over 2} \rangle \, e^{ipy}   \, dy  } \nonumber \\
 &=& {1\over 2\pi^2} \int \langle x- {y\over 2} | -x - {y\over 2} \rangle \, e^{ipy} \, dy,
 \nonumber \\
 &=& {\delta(2x)\over 2\pi^2} \int  e^{ipy} \, dy,
 \nonumber \\
 &=& {\delta(x) \delta(p)\over 2\pi},
\end{eqnarray}
where the first equality is obtained by using the evolution equation and we have used the identity $\int  e^{ipy} \, dy = 2\pi \, \delta(p)$. Hence, using the overlap formula (\ref{eq:overlapformula}), we get
\begin{eqnarray}  \label{eq:parity-wigner-at-origin}
\frac{1}{\pi}\tr \left( \hat \rho \, (-1)^{\hat n} \right)
 &=&  2\pi \int W(x,p) \, {\delta(x) \delta(p)\over 2\pi}\, dx \, dp ,
 \nonumber \\
 &=& W(0,0),
\end{eqnarray}
which proves the well-known fact that the expectation value of the photon number parity is proportional to the value of the Wigner function at the origin. This fact has already been used to reconstruct a quantum state using unbalanced homodyne detection in Refs. \cite{wallentowitz_unbalanced_1996,banaszek_direct_1996}.


In order to express the QCS in terms of the Wigner function of  state $\hat \rho_d$ (see Sec. \ref{sec:phasespace}), we also need to compute the Weyl transform of the operator $({\hat x}^2 + {\hat p}^2) \, (-1)^{\hat n} /\pi$ and then apply the overlap formula. We first calculate the Weyl transform of ${\hat x}^2 \, (-1)^{\hat n} /\pi$, namely
\begin{eqnarray}
\lefteqn{ {1\over 2\pi} \int \langle x- {y\over 2} | {{\hat x}^2 \,(-1)^{\hat n} \over \pi}| x + {y\over 2} \rangle \, e^{ipy}   \, dy  } \nonumber \\
 &=& {1\over 2\pi^2} \int \left(x - {y\over 2}\right)^2 \langle x- {y\over 2} | -x - {y\over 2} \rangle \, e^{ipy} \, dy,
 \nonumber \\
 &=& {\delta(2x)\over 8\pi^2} \int  y^2 \, e^{ipy} \, dy,
 \nonumber \\ 
 &=& -{\delta(x)\, \delta''(p)\over 8\pi} ,
\end{eqnarray}
where we have used the identity $\int  y^2 \, e^{ipy} \, dy = - 2\pi \, \delta''(p)$. 
Using the overlap formula, we then obtain
\begin{eqnarray}
\frac{1}{\pi}\tr \left( \hat \rho \, {\hat x}^2 \,(-1)^{\hat n} \right)
 &=&  - 2\pi \int W(x,p) \, {\delta(x) \, \delta''(p)\over 8\pi}\, dx \, dp ,
 \nonumber \\
 &=& - {1\over 4} {\partial^2 W \over \partial p^2}\bigg|_{x=0,p=0}  ,
\end{eqnarray}
where we have used the identity
$\int f(x) \, \delta''(x) \ dx = f''(0)$. Similarly, we have
\begin{eqnarray}
\frac{1}{\pi}\tr \left( \hat \rho \, {\hat p}^2 \,(-1)^{\hat n} \right)
= - {1\over 4} {\partial^2 W \over \partial x^2}\bigg|_{x=0,p=0} ,
\end{eqnarray}
so that
\begin{eqnarray}
\label{eq:laplacian-wigner-at-origin}
\frac{1}{\pi} \tr \left( \hat \rho  \, (\hat{x}^2 + \hat{p}^2)\, (-1)^{\hat n} \right) =  - {1 \over 4} \Delta W\bigg|_{x=0,p=0},
\end{eqnarray}
where $\Delta$ stands for the Laplacian.

\section{Gaussian purity, overlap, and QCS}
\label{sec:appendixGaussian}



Let us show how the known formulas of the purity, overlap, and QCS of Gaussian states can be painlessly rederived from the phase-space expressions for these quantities that we obtained in Sec~\ref{sec:phasespace}. To do this, we start from the expression of the Wigner function of a Gaussian state centered at origin, namely,  
\begin{equation}
\label{eq:wignerGaussian}
W(x,p)=\frac{1}{2\pi \sqrt{\det\gamma}}   \exp ^{\left(- \frac{1}{2} \, \mathbf{r}^T \gamma^{-1} \, \mathbf{r}  \right)},
\end{equation}
with $\mathbf{r}=(x,p)^T$ and where $\gamma$ is the covariance matrix
\begin{equation}
    \gamma = \begin{pmatrix}
    \sigma_x^2 & \sigma_{xp} \\
    \sigma_{xp} & \ \sigma_p^2 
    \end{pmatrix}.
\end{equation}
Here, $\sigma_x^2$ and $\sigma_p^2$ are the variances of the $x$ and $p$ quadratures, respectively, and $\sigma_{xp}$ stands for the covariance. It is sufficient to consider a centered state here since the purity and QCS are both invariant under displacements in phase-space (and displacements are easy to account for in the overlap between two states).

\subsection{Purity}


We can recover the expression of the purity by using Eq.~\eqref{eq:overlapWd}.
If $\hat \rho$ is a Gaussian state centered at origin, then $\hat \rho_d=\hat \rho$ because the product of two identical Gaussian states impinging on a beam splitter remains unchanged~\cite{WeedbrookRMP}. 
Thus, $W_d(0,0)= 1/(2\pi \sqrt{\det \gamma})$ according to Eq. \eqref{eq:wignerGaussian}, so that we have, from Eq.~\eqref{eq:overlapWd},
\begin{eqnarray}
\tr \left( \hat \rho^2\right) 
= \pi \, W_d(0,0) = \frac{1}{2 \sqrt{\det \gamma}},
\end{eqnarray}
which is indeed the usual formula for the purity of a Gaussian state  \cite{WeedbrookRMP}. 

Interestingly enough, it then follows from Eq.~\eqref{eq:purity-bis} that, for a centered Gaussian state, 
$$
\mathrm{Tr} \left( \hat \rho^2 \right)= \mathrm{Tr}\left( \hat \rho \, (-1)^{\hat n} \right).
$$

\subsection{Overlap}

We now use Eq.~\eqref{eq:overlapWd} in the special case where $\hat\rho_a$ and $\hat\rho_b$ are two Gaussian states (both assumed to be centered for simplicity). 
First note that, if $\hat\rho_a$ ($\hat\rho_b$) is characterized by the covariance matrix $\gamma_a$ ($\gamma_b$), then the output state $\hat\rho_d$ of the 50:50 beam splitter is a centered Gaussian state with covariance matrix $\gamma_d = (\gamma_a+\gamma_b)/2$ \cite{WeedbrookRMP}. 
Its Wigner function as given by Eq. (\ref{eq:wignerGaussian}) admits the value at origin
\begin{equation}
W_d(0,0)=\frac{1}{2\pi \sqrt{\det \gamma_d}} =\frac{1}{\pi \sqrt{\det(\gamma_a+\gamma_b)}} .
\end{equation}
Using Eq.~\eqref{eq:overlapWd}, we then find
\begin{equation}
\tr(\hat\rho_a \, \hat\rho_b) =
\frac{1}{\sqrt{\det(\gamma_a+\gamma_b)}},
\end{equation}
which is the well-known formula  for the overlap between two centered Gaussian  states  \cite{marian_uhlmann_2012}.



\bigskip
\subsection{Quadrature coherence scale}

If the input state is Gaussian, then $W_d$ is simply equal to the Wigner function of the input state and has the form of Eq. (\ref{eq:wignerGaussian}). In that case, the Laplacian at origin is easily expressed as:
\begin{eqnarray}
\Delta W_d\big|_{x=0,p=0} = - \frac  {\tr (\gamma^{-1})}{2 \pi \sqrt{\det \gamma}},
\end{eqnarray}
while $W_d(0,0)= 1/(2\pi \sqrt{\det \gamma})$, so that we conclude that 
the QCS of a Gaussian state of covariance matrix $\gamma$ is
\begin{eqnarray}
\mathcal C^2(\hat \rho)  
= - {1\over 4} {\Delta W_d \over W_d}\bigg|_{x=0,p=0} = {1\over 4} \,  \tr(\gamma^{-1}),
\end{eqnarray}
in agreement with the expression proven in Ref.~\cite{hertz_quadrature_2020}.

\bibliography{main.bib}

\end{document}